\documentstyle[12pt]{article}
\renewcommand{\section}[1]{\addtocounter{section}{1}
\vglue 0.5cm
{\noindent{\large\bf \thesection ~#1}}
\vglue 0.4cm }
\renewenvironment{thebibliography}[1]
 { \begin{list}{\arabic{enumi}.}
    {\usecounter{enumi} \setlength{\parsep}{0pt}
     \setlength{\itemsep}{3pt} \settowidth{\labelwidth}{#1.}
     \sloppy
    }}{\end{list}}
\begin{document}
\newcommand{\eqn}[1]{(\ref{#1})}
\newcommand{\ft}[2]{{\textstyle\frac{#1}{#2}}}
\newcommand{\dr}{\raise.3ex\hbox{$\stackrel{\leftarrow}{\partial}$}{}}
\newcommand{\drsmall}
    {\raise.1ex\hbox{$\scriptstyle\stackrel{\leftarrow}{\partial}$}{}}
\newcommand{\dl}{\raise.3ex\hbox{$\stackrel{\rightarrow}{\partial}$}{}}
\newcommand{\rh}{\right\rangle}
\newcommand{\lh}{\left\langle}
\newcommand{\afn}{a\!f\!n}
\newcommand{\PV}[1]{\underline{#1}}
\newcommand{\pl}{(\partial X^\mu )}
\newcommand{\plb}{(\partial X^\nu )}
\newcommand{\plg}{(\partial X^\rho )}
\newcommand{\plr}{(\partial X^\sigma )}
\newcommand{\pls}{(\partial X^\tau )}
\newcommand{\bpl}{(\bar{\partial }X^\mu )}
\newcommand{\ds}{d_{\mu \nu \rho }}
\newcommand{\na}{\nabla }
\newcommand{\dkt}{\delta _{KT}}
\newcommand{\dtx}[1]{\frac{\tilde\partial}{\partial\xi^{#1}}}
\renewcommand{\theequation}{\arabic{equation}}
\newsavebox{\uuunit}
\sbox{\uuunit}
    {\setlength{\unitlength}{0.825em}
     \begin{picture}(0.6,0.7)
        \thinlines
        \put(0,0){\line(1,0){0.5}}
        \put(0.15,0){\line(0,1){0.7}}
        \put(0.35,0){\line(0,1){0.8}}
       \multiput(0.3,0.8)(-0.04,-0.02){12}{\rule{0.5pt}{0.5pt}}
     \end {picture}}
\newcommand {\unity}{\mathord{\!\usebox{\uuunit}}}
\newcommand{\delb}{\bar{\partial}}
\begin{titlepage}
\begin{flushright} KUL-TF-94/27\\ hep-th/9410162  \\
                   October 1994\\
\end{flushright}

\vfill
\begin{center}
{\large\bf  Regularisation, the BV method,  and\\ the antibracket
cohomology $^1$
}\\ \vskip 27.mm
{\bf  Walter Troost $^2$ and Antoine Van Proeyen $^{3}$
}\\ \vskip 1cm
Instituut voor Theoretische Fysica
        \\Katholieke Universiteit Leuven
        \\Celestijnenlaan 200D
        \\B--3001 Leuven, Belgium\\[0.3cm]
\end{center}
\vfill
\begin{center}
{\bf Abstract}
\end{center}
\begin{quote}
\small
We review the Lagrangian Batalin--Vilkovisky method for gauge theories.
This includes gauge fixing, quantisation and regularisation.
We emphasize the role of cohomology of the antibracket
operation. Our main example is $d=2$ gravity, for which we also
discuss the solutions for the cohomology in the space of local
integrals. This leads to the most general form for the action,
for anomalies and for background charges.

\vspace{2mm} \vfill \hrule width 3.cm
{\footnotesize
\noindent $^1$ presented at the G\"ursey memorial conference I
                   on strings and symmetries
              Bogazici University, Istanbul, Turkey
                       6-10 June 1994. To be published by Springer
Verlag in the series "Lecture Notes on Physics". \\
\noindent $^2$ Onderzoeksleider, NFWO, Belgium;
E--mail : Walter\%tf\%fys@cc3.kuleuven.ac.be \\
E--mail :~Walter.Troost@fys.kuleuven.ac.be.\\
\noindent $^3$ Onderzoeksleider, NFWO, Belgium;
E--mail :~Antoine.VanProeyen@fys.kuleuven.ac.be.
 }
\normalsize
\end{quote}
\end{titlepage}
\section{Introduction.}
The Batalin--Vilkovisky method for quantisation of
general gauge theories has been introduced in \cite{BV}.
It is applicable to all known gauge theories. Another advantage of the
method is that it gives a comprehensive picture of the gauge
fixing procedure. It uses a space of fields and antifields endowed
with a symplectic structure defining an antibracket operation. The
latter is a Poisson--like structure, and field transformations which
are canonical with respect to this antibracket play an important part.
Gauge fixing is essentially such a canonical transformation. We will
summarize the essential ingredients and tools. In comparison with
previous lectures \cite{bvsb,bvberk} we will put more emphasis here on
the role of cohomology. We will use $d=2$ gravity as our main example, and
refer to \cite{bvberk} for chiral $W_3$.

In section 2 we will recall some definitions, introduce antifields,
antibrackets and the `extended classical action'. In the third section we
introduce three different cohomologies and indicate the different roles
they play. In section 4 we will review how gauge fixing is obtained by a
canonical transformation. Then we will turn to the quantum theory.
General principles will be given in section 5, where also the
regularisation (based on Pauli--Villars) will be introduced
\cite{anomPV,anombv}. Anomalies are discussed in section 6, and illustrated
with a
calculation for $d=2$ gravity. The latter will be treated also from a
purely cohomological point of view in section 7. That section reports
work in collaboration with F. Brandt \cite{anom2dgr}.
Besides the most general form for the classical action and for
anomalies, we also obtain the most general form for background charges.

This lecture is not self-contained. The reader is urged to have
also \cite{bvsb,bvberk} at hand, as we will refer to specific formulas
in these reviews. The relevant equations and sections will be referred to
as (I.*), resp. (II.*). A more complete text is still in preparation
\cite{BVboek}. Many examples can also be found in \cite{QJS}.
\pagebreak[0]

\section{The ingredients}
Let us start by introducing the setup for $d=2$ gravity with scalar matter.
The classical fields are
$
\phi ^i=\{ X^\mu ,g_{\alpha\beta}\}
$,
where $X^\mu $ are the $D$ scalar fields, and $g_{\alpha\beta}$ is
the two dimensional metric. There are two types of gauge symmetries,
for which we introduce ghosts $ c^a=\left\{ \xi^\alpha, c\right\} $,
namely $ \xi^\alpha $ for
the general coordinate transformations ($\alpha=+$ or $-$) and $c$
for the local dilatations.
Note that we do not introduce gauge fields for these transformations.
For the diffeomorphisms they are nevertheless effectively included in the
metric $g_{\alpha\beta}$, and in fact one could equally set up this same
problem using the zweibeins instead of the metric. For the dilatations
however they remain excluded.

We denote all these fields collectively as $\Phi^A= \{\phi^i,c^a\}$.
We introduce antifields for all of them, $\Phi^*_A$. They have opposite
statistics. Then one defines an extended action $S(\Phi,\Phi^*)$,
function of all the Fields\footnote{We need a common denomination for
talking about  fields and antifields together. For this we will use
`Fields' with a capital F. We will use the notation
$z^\alpha =\{\Phi ^A, \Phi ^*_A\}$ (with $A=1,..., N$, and
$\alpha =1, ..., 2N$). }.
The construction of this extended action
was explained in detail in \cite{bvsb,bvberk}, and leads
to\footnote{We omit
$\int d^2x$ for actions and anomaly expressions, here and in the sequel.
(Anti-)Symmetrisation of indices is defined by
 $f_{(\mu\nu)}=\ft 12(f_{\mu\nu}+f_{\nu\mu})$, etc.}
\begin{eqnarray} S&=&
-\ft12\sqrt{g}\, g^{\alpha\beta}\partial_\alpha X^\mu \cdot
\partial_\beta X_\mu \nonumber\\ &&
\mbox{}+X^*_\mu \xi^\alpha \partial_\alpha X^\mu  
+ g^{*\alpha \beta} \left( \xi^\gamma\partial_\gamma g_{\alpha
\beta}+2g_{\gamma (\alpha}\partial_{\beta )}\xi^\gamma  + c
g_{\alpha\beta} \right)\nonumber\\
&&\mbox{} -\xi_\beta^*  \xi^\alpha \partial_\alpha \xi^\beta  -
c^* \xi^\alpha \partial_\alpha c\ .  \label{S2dgr}
 \end{eqnarray}
The first line is the classical action,
and the other lines are the BRST
transformations of the fields multiplied by their corresponding
antifields. In general, more terms may be present, for example expressing
relations in an open gauge algebra. Introducing now
an {\bf antibracket}
\begin{equation}
(F, G) = F\frac{\dr }{\partial \Phi ^A}\cdot \frac{\dl}{\partial
\Phi ^*_A} G - F\frac{\dr}{\partial \Phi ^*_A}\cdot\frac{\dl
}{\partial \Phi ^A} G\ , \label{abracket}
\end{equation}
It is clear that the extended action satisfies
\begin{equation} (S,S)=0 \ .\end{equation}
When one checks the terms in this equation with a specific type of
antifields, one will notice that it expresses the gauge invariance of
the classical action, as well as the commutator algebra of
the gauge transformations and their Jacobi identities. In general,
for example with open gauge algebras,
it will also include the other defining relations of the gauge theory.

One assigns ghost numbers 0 for the classical fields $\phi^i$, 1 for the
ghosts $c^a$, and for the antifields a ghost number such that the sum of
the ghost number of a field and its corresponding antifield always adds
up to $-1$. As a consequence,by \eqn{abracket}, this gives a ghost number
assignment to $(F, G)$ which is 1 lower than that of $FG$.

We invite the reader now to look to the example of $W_3$ in section
II.1 and II.2. This example was treated in full using the BV method
in \cite{anomw3}.
\pagebreak[1]

Let us summarize the essential properties of the extended classical
action.
It is a function $S(z)$ of all the Fields $z$.
\begin{description}
\item[Classical limit :]
  $S(\Phi , \Phi ^*=0)$ is the classical action.
\item[Master equation :] $(S,S)=0$, which implies that\footnote{The argument
was
repeated in section I.3.1.}
$R_{\alpha \beta }\equiv {\partial \over\partial z^{\phantom{\beta
}\!\!\!\alpha} } {\partial \over\partial z^\beta }S$ has rank $\leq N$.
\item[Properness condition :] the rank of  $R$ is equal to $N$. This
property essentially means that we have to include all
symmetries using ghosts, and also all zero modes. It will
allow us to define the path integral.
\end{description}
It has been proven in several steps \cite{openalg,Henn,anomw3}
that there is a local solution to these conditions for all classical
actions (under certain regularity conditions satisfied by all
reasonable gauge theories).
The proof is very long, and will not be repeated here.
The main tool
is the `Koszul--Tate' differential, which we will define below.
\section{Cohomologies}
We now introduce several different but related
differentials and cohomologies. From the extended action one defines
the nilpotent operator
\begin{equation}
{\cal S} F(z)\equiv  (S,F) \ .
\end{equation}
The nilpotency follows from the Jacobi identity of antibrackets and
the master equation:
\begin{equation}
{\cal S}^2F=(S,(S,F))=\ft12((S,S),F)=0 \ .
\end{equation}
The operator ${\cal S}$ raises the ghost number by 1,
i.e.: $gh ({\cal S}F) = gh(F)+1$.
The cohomology of this operator in the space of
local functions of ghost number 0 gives physically meaningful quantities.
They are formed by arbitrary gauge invariant functionals, and are defined
up to field equations.

We now introduce a second grading: the antifield number ($\afn$).
This is $-gh$ for the Fields of negative
ghost number (antifields in the `classical basis', the one used so far),
and 0 for Fields of non--negative ghost
number.  Check that the
extended action \eqn{S2dgr} has zero ghost number, but the 3 lines
have antifield numbers 0,1 and 2 respectively.
The action of ${\cal S}$ can be split according to its change in
the antifield number:
\begin{equation} \begin{array} {ccccccccccc}
{\cal S}&=& \dkt &+&\Omega& +& D_1 &+&D_2 &+& \ldots\\
\afn &      & -1   &&  0    &  & 1 &  & 2
\end{array}            \label{expcalS}
\end{equation}
Observe that the antifield number at most diminishes by one. This
part defines the Koszul--Tate (KT) differential
(introduced in BV quantisation in \cite{Henn}). Explicitly, the main
properties are:
\begin{equation} \begin{array}{ll}
\dkt \Phi = 0                             & \mbox{vanishes on fields}   \\
\dkt \phi^*_i = \frac{S^0 \drsmall}{\partial \phi^i}
                                          & \mbox{field equations} \\
\dkt c^*_a = \phi^*_i R^i{}_a &
\end{array} \end{equation}
($S^0$ is the action without antifields, and $R^i{}_a$ determines its
gauge transformations). Now the nilpotency property
${\cal S} ^2 = 0 $ implies
\begin{equation}
\begin{array}{ll}
\dkt^2 = 0 & \mbox{KT is nilpotent }\\[3mm]
\Omega\dkt +\dkt \Omega =0 & \\[3mm]
\Omega^2 = -\dkt D_1 - D_1\dkt& \mbox{On--shell nilpotent BRST}
\end{array} \label{expcalS0}
\end{equation}
The equation \eqn{expcalS} also defines $\Omega$, which will be
called the BRST operator. Its action on functionals of
the fields is given by
\begin{equation}
 \Omega\Phi= \left.{\cal S}\Phi \right|_{\Phi^*=0}  \label{defOmf}
\end{equation}
We will come back to the meaning of the last lines of \eqn{expcalS0}
in a moment.

We have now {\em 3 different cohomologies}, related to ${\cal S}$,
$\dkt$ and $\Omega $.
We should also be careful to
distinguish cohomology in the space of local functions, or in the
space of integrals thereof (this is cohomology $mod\ d$, denoted
$H^g ({\cal S}\mid d)$, where $d$ is the space--time differential).

1. $\dkt$ acts in the space of functions of $\Phi^*_A$ and $ \phi^i $, i.e.
functions of ghost number negative or zero: it vanishes on the
ghosts, and also the image of $\dkt$ on the
antifields or classical fields does not contain the ghosts.
As a grading one uses the antifield number, and
$H^k(\dkt)$ then denotes the cohomology of $\dkt$ in local
functions of antifield number $k$. Its main property is that
\begin{equation}
H^k (\dkt ) = \delta^k_0\times
 \mbox{ functions on the stationary surface},
\end{equation}
where the stationary surface refers to the surface in the space of
fields of ghost number 0 (`classical fields' $\phi^i$) where the
classical field equations are satisfied\footnote{To be more exact,
in some cases one should define this concept using the concept of
functions `not proportional to field equations', see the discussion
on `evanescent functions' in section 3.1 of \cite{anomw3}.}.
Also for integrals over local functions of ghost number zero, or
$H^0 (\dkt \mid d)$, such a property holds. But there is not such a
general statement for $H^k (\dkt \mid d)$ with $k>0$ (in \cite{bbh}
some general results have been given, relating e.g. the $k=1$ case
to rigid symmetries).

2. The last line of \eqn{expcalS0} implies that $\Omega$ is nilpotent
in the cohomology space of $\dkt$. One can define a cohomology
of $\Omega$ in the cohomology space $H^k (\dkt )$. This is the BRST
cohomology, which therefore is defined {\bf on the stationary surface}.
So a function $F$ is in the cohomology of $\Omega$, and nontrivial, if
\begin{equation}
\Omega F\approx 0\qquad\mbox{ and }\qquad F\not\approx \Omega G\ ,
\end{equation}
where $\approx$ means equality on the stationary
surface, defined with the classical field equations,
and $F$ satisfies $\dkt F=0$.
This is denoted by $H^k\left(\Omega \mid
H(\dkt)\right)$. The grading of $\Omega$ is the ghost number.
Therefore $k$ refers here to the ghost number. For $k=0$ the
cohomology of local functions now identifies functions which differ
by gauge transformations, and, combined with the result for $\dkt$,
this implies that we are left with the physical observables.

3. ${\cal S}$ acts on fields and antifields.
Its grading is the ghost number.
The cohomology of ${\cal S}$ combines the results of the previous
two cohomologies:
\begin{eqnarray}
\mbox{for }k<0\ : && H^k({\cal S})=  0 \nonumber\\
\mbox{for }k\geq 0\ : && H^k({\cal S})= H^k(\Omega ) \equiv
H^k\left(\Omega \mid H(\dkt)\right)\ .
\end{eqnarray}
Note that this statement for $k\geq 0$ applies also for cohomology
$mod\ d$ (local integrals).  It is important that to define the
cohomologies with the
BRST method we have to use `weak equalities' (even for a `closed
algebra'), while in the antibracket cohomology the inclusion of the
antifields allows us to work with unqualified equalities.

\section{Gauge fixing }
Let us illustrate the gauge fixing with a simple example. Consider
the theory with classical fields $\phi ^i=\{ X^\mu ,h\}$ and
classical action
\begin{equation} S_0 = \partial X^\mu \bar\partial X_\mu \ .\end{equation}
Trivially, $h$ does not occur in it,
which we can interpret as a gauge invariance
$\delta h= \epsilon $. The extended action is
\begin{equation}
S=  \partial X^\mu \bar\partial X_\mu+ h^* c \ .
\end{equation}
For propagators, one wishes to invert the matrix of second derivatives
of the action w.r.t. fields, but it is singular.
Consider however the matrix of second derivatives w.r.t. Fields,
$\partial_\alpha\partial_\beta S$, and make a slight rearrangement:
\begin{equation}
\begin{array}{lcr}
 \begin{array}{lc}
   & \begin{array}{cccccc}
       X& h& c & X^* & h^* & c^*
     \end{array}  \\
     \begin{array}{l}
      X^\mu\\ h\\ c \\ X^*_\mu \\ h^* \\ c^*
     \end{array} &
     \pmatrix{*&0 & 0 & 0 & 0 & 0 \cr 0 &0 & 0 & 0 & 0 & 0\cr
              0 &0 & 0 & 0 & * & 0\cr 0 &0 & 0 & 0 & 0 & 0\cr
              0 &0 & * & 0 & 0 & 0\cr 0 &0 & 0 & 0 & 0 & 0\cr}
 \end{array}
&\rightarrow&
 \begin{array}{lc}
   & \begin{array}{cccccc}
       X& h^*& c & X^* & h & c^*
     \end{array}  \\
     \begin{array}{l}X^\mu\\ h^*\\ c \\ X^*_\mu \\ h \\ c^*
     \end{array} &
     \pmatrix{*&0 & 0 & 0 & 0 & 0 \cr 0 &0 & * & 0 & 0 & 0\cr
              0 &* & 0 & 0 & 0 & 0\cr 0 &0 & 0 & 0 & 0 & 0\cr
              0 &0 & 0 & 0 & 0 & 0\cr 0 &0 & 0 & 0 & 0 & 0\cr}
 \end{array}
\end{array}.
\end{equation}
The stars in the matrix denote the non--zero entries. The rank of
the matrix is indeed half its dimension (properness condition).
The invertible part is just the left upper corner of the second matrix.
(The other entries happen to be zero because of the simplicity of the
model). If we reassign the antifield $h^*$ to be a field,
renaming it $b$, then
$h$ becomes an antifield $b^*$, and the action in terms of the
new fields has no zero--modes. This is what we call gauge--fixed.

In general {\bf gauge fixing is a canonical transformation such that the
new `fields' have no zero--modes.} Later we will
choose these fields as integration variables of the path integral.

Also in $d=2$ gravity gauge fixing is accomplished by just such a change
of name. We define
\begin{equation}
g_{\alpha\beta}=\eta_{\alpha\beta}  +b^*_{\alpha\beta}\ ;\qquad
g^{*\alpha\beta}=
- b^{\alpha\beta} \ .  \label{gf2dgr}
\end{equation}
We can now consider two bases:
\begin{equation}
\begin{array}{l|ccccccccc}
\mbox{classical basis}& X^\mu & g_{\alpha\beta} & \xi^\alpha & c & X^*_\mu &
g^{*\alpha\beta} & \xi^*_\alpha & c^*
\\
statistics   &+     & +   & - & -& -&-&+&+\\
ghost\ number &0     & 0 & 1 & 1&-1& -1& -2  & -2  \\
\mbox{gauge fixed basis}& X^\mu & b^*_{\alpha\beta}  & \xi^\alpha & c
& X^*_\mu & b^{\alpha\beta}  &
\xi^*_\alpha & c^*  \end{array}
\end{equation}
It is now easy to see that, substituting \eqn{gf2dgr} in
\eqn{S2dgr}, the part of the extended action which is independent of
antifields in the gauge--fixed basis has no gauge invariances any
more, i.e. it is properly called gauge fixed.
Note that $b^{\alpha\beta}$ plays the role of the antighost.

There are other types of canonical transformations that are often
important. The general type of a canonical transformation
$\{\Phi,\Phi^*\}$ to $\{\tilde\Phi, \tilde\Phi^*\}$ where the field--field
part is invertible (this does not include the above transformation
\eqn{gf2dgr}), can be  generated from a (fermionic) function
$F(\Phi , \tilde\Phi^*)$ :
\begin{equation}
\tilde\Phi^A=\frac{\partial F(\Phi , \tilde\Phi^*)
}{\partial \tilde\Phi^*_A}\hspace{2cm}
\Phi^*_A=\frac{\partial F(\Phi , \tilde\Phi^*)}{\partial \Phi ^A}\ .
\label{cantrF}\end{equation}
We will here illustrate the use of such a transformation,
not for gauge fixing (that will be done below), but to achieve
a simplification in the extended
action of $d=2$ gravity. Consider the transformation from
$\{\Phi\}=\{ X^\mu , g_{\alpha\beta}, \xi^\alpha ,c\}$ and their
antifields to
$\{\tilde\Phi\}= \{\tilde X^\mu, h_{++},h_{--},e,c^\alpha,\tilde c\}$
and the corresponding antifields generated by
(an integral over $d^2x $ is implied)
\begin{eqnarray}
F&=&\tilde X^*_\mu X^\mu
+e^*\sqrt{g}
+\tilde c^*\left( c\sqrt{g}
 +\partial_\alpha \xi^\alpha \sqrt{g}
 \right)\nonumber\\ &&
 +h^{++\,*}
\frac{g_{++}}{g_{+-}+\sqrt{g}}
 +h^{--\,*}
 \frac{g_{--}}{g_{+-}+\sqrt{g}}  \nonumber\\
&&+c_+^* \left( \xi^+ +
 \frac{g_{--}}{g_{+-}+\sqrt{g}}\xi^-
\right)
+c_-^* \left( \xi^- +
 \frac{g_{++}}{g_{+-}+\sqrt{g}}\xi^+
\right)  \ . \label{Fsimpl}
\end{eqnarray}
This leads to
\begin{eqnarray}
S&=&\frac{1}{1-h_{++}h_{--}}\left(-\nabla _+ X^\mu \cdot
\nabla _- X^\nu \eta _{\mu \nu } + X^*_\mu
 c^\alpha \nabla_\alpha X^\mu \right) \nonumber\\
&&+ h^{++\,*}\nabla _+ c^- + h^{--\,*}\nabla _- c^+
+e^*\tilde c \nonumber\\ && - c^*_+ c^+\partial _+ c^+
- c^*_- c^-\partial _- c^-\ ,
\end{eqnarray}
where  $ \nabla _+\equiv \partial _+  -h_{++}\partial
_- +\lambda (\partial _-h_{++}\cdot ) $ ($\lambda$ is the spin:
lower $+$ indices, or upper $-$ indices ...).
Here the term
$e^* \tilde c$ exhibits {\bf a trivial system}: ${\cal S}e = \tilde c$ and
${\cal S}\tilde c^*= e^*$. The cohomology in the quartet
$\{ e,\tilde c,e^*,\tilde c^*\}$ is trivial, it can therefore be omitted.
The reader may also consult section I.3.2 for a review on the
applications of canonical transformations, and section I.3.4 on the
trivial systems.
In the new basis gauge fixing is again performed just by the
transformation $h^{++\,*}= b^{++}$ and $h_{++}=-b^*_{++}$.

The canonical transformation of the form \eqn{cantrF} is often used
for gauge fixing. Consider $F$ of the particular form
\begin{equation}
F=\Phi^A\Phi^*_A +\Psi(\Phi)\ .  \label{cantranF}
\end{equation}
The first term by itself produces just the identical transformation. The
second is the so--called gauge fermion $\Psi$,
which was needed in \cite{BV}, although it should be clear now that
this method is a particular case. For Maxwell
theory for example, the minimal extended action would be just
$ S=-\frac{1}{4}F_{\mu \nu }F^{\mu \nu }+A^*_\mu \partial ^\mu c$. One
can first add an auxiliary field term $-\frac{1}{2}b^*b^*$. In the
cohomology language this is just again a trivial system
${\cal S}b = b^*$ and $ {\cal S}b^* = 0$. Then the canonical
transformation we need for gauge fixing is of the type (\ref{cantranF})
with
$\Psi = b\partial _\mu A^\mu $. The reader can check that this gives
the usual gauge--fixed action. Note that, although we had to
introduce a trivial system before the canonical transformation, even
in this simple example there is still a simplification w.r.t. the
usual BRST procedure. In BRST one first introduces also an auxiliary
field $\lambda$. This is necessary to have a nilpotent
BRST--operator off shell.
However the algebra of ${\cal S}$ is always closed.

We have related the physical observables to the cohomology of
the operator ${\cal S}$.
The essential principle is that we always use canonical
transformations, and add or delete trivial systems.
A canonical transformation does not change the cohomology
of this operator, and neither do the trivial systems.
In a gauge--fixed basis we
may again define a BRST operator by \eqn{defOmf}, where the fields
and antifields are now those of the gauge--fixed basis. Then one can
prove that the cohomology of $\cal S$ is equivalent to the `weak'
cohomology of $\Omega$ (using the new field equations of all the
fields). This
proves that the BRST cohomology gives the physical observables again, but
also shows the advantage of the ${\cal S}$--cohomology: it includes
the information in the  field equations, as
${\cal S} \Phi^* = \frac{S\drsmall}{\partial\Phi}$.
\section{Quantum theory and Pauli--Villars regularisation}
The general principles of the quantum treatment in the BV framework
have been described in section~II.4, to which we refer for an exposition
of our method.
The advantage of the Pauli--Villars (PV) regularisation is that it has a
Lagrangian formulation that interfaces nicely with the BV framework.
We will now describe some additional aspects of the PV
procedure, using $d=2$ gravity as an illustration.\footnote{The
notation used here and in the sequel
differs slightly from the one used in (I) and (II),
viz. we denote the PV--partner of a Field by  underscoring its symbol.}
The quantum theory, in one loop approximation, is based on the action
\begin{equation}
S_T= S(z) + S_{PV}(z,\PV{z}) + S_M(z,\PV{\Phi})\ .
\end{equation}
In this formula, the massless PV extended action and the mass terms
for the PV fields are  given by
\begin{equation}
 S_{PV} =\frac 12 \underline{z}^\alpha \left( \frac{\dl}{\partial
z^{\phantom{\beta }\!\!\!\alpha}
}S \frac{\dr}{\partial z^\beta } \right) \underline{z}^\beta \
\qquad\mbox{and}\qquad
S_M= -\frac 12 M^2 \underline{\Phi} ^A T_{AB}(z) \underline{\Phi} ^B\ ,
\label{SPV}\end{equation}
where $T_{AB}(z)$ is largely arbitrary but must be invertible.
Note that this last condition implies
that, formally,   in the limit $M\rightarrow\infty $,
the PV Fields are cohomologically trivial since
\begin{equation}
(S,\underline{\Phi}^*_A)=M^2 T_{AB}\underline{\Phi}^B\ .
\end{equation}

In the example of $d=2$ gravity, we get from \eqn{S2dgr}
\begin{eqnarray}
S_{PV}&=&
  -\ft12\sqrt{g} g^{\alpha\beta}\partial_\alpha \PV{X}^\mu \cdot
  \partial_\beta \PV{X}_\mu\nonumber\\
&& +\PV{g}_{\gamma\delta} \sqrt{g}\left( g^{\alpha\gamma}g^{\beta\delta}
-\ft12
g^{\alpha\beta}g^{\gamma\delta}\right) \partial_\alpha \PV{X}^\mu \cdot
  \partial_\beta {X}_\mu\\
&& -\ft12\PV g _{\gamma\delta}\PV g_{\epsilon\phi}\sqrt{g}\left(
g^{\alpha\gamma}
g^{\delta\epsilon}g^{\phi\beta}-\mbox{traces}\right)
\partial_\alpha X^\mu \cdot
  \partial_\beta {X}_\mu\nonumber\\
&& +\PV{X}^*_\mu \xi^\alpha \partial_\alpha \PV{X}^\mu
          +\PV{X}^*_\mu \PV{\xi}^\alpha \partial_\alpha {X}^\mu
          +X^*_\mu \PV{\xi}^\alpha \partial_\alpha \PV{X}^\mu
          \nonumber\\ &&
          +\PV{g}^{*\alpha \beta}\left(
             \xi^\gamma\partial_\gamma \PV{g}_{\alpha\beta}
           +2\PV{g}_{\gamma(\alpha}\partial_{\beta)}\xi^\gamma
           +c \PV{g}_{\alpha\beta}\right)
           \nonumber\\ &&
           +\PV{g}^{*\alpha \beta}\left(
             \PV{\xi}^\gamma\partial_\gamma g_{\alpha\beta}
             +2g_{\gamma(\alpha}\partial_{\beta)}\PV{\xi}^\gamma
             +\PV{c} g_{\alpha\beta}\right)
             +g^{*\alpha \beta}\left(
             \PV{\xi}^\gamma\partial_\gamma \PV{g}_{\alpha\beta}
             +2\PV{g}_{\gamma(\alpha}\partial_{\beta)}\PV{\xi}^\gamma
             +\PV{c} \PV{g}_{\alpha\beta}\right)
\nonumber\\
&& -\PV{\xi}_\beta^*  \PV{\xi}^\alpha \partial_\alpha \xi^\beta
          -\PV{\xi}_\beta^*  \xi^\alpha \partial_\alpha      \PV{\xi}^\beta
          -     \xi_\beta^*  \PV{\xi}^\alpha \partial_\alpha \PV{\xi}^\beta
 - \PV{c}^* \PV{\xi}^\alpha \partial_\alpha c
          -      c^* \PV{\xi}^\alpha \partial_\alpha \PV{c}
          - \PV{c}^*      \xi^\alpha \partial_\alpha \PV{c}\ .
\end{eqnarray}
Gauge fixing is now performed by the canonical transformation \eqn{gf2dgr}
together with
\begin{equation}
\PV{g}^{*\alpha \beta}=-\PV{b} ^{ \alpha \beta}  \ ;\qquad
  \PV{g}_{ \alpha \beta}=\ \PV{b} ^*_{\alpha \beta} \ .
  \end{equation}
Note that here we introduced the PV system already before gauge fixing.
To keep the correspondence between the PV action and the action of
the ordinary fields as in \eqn{SPV}, their  canonical
transformations must be related. This question is treated in more detail
in (I.42).

The PV mass term should be quadratic in the PV partners of the fields in
the gauge--fixed basis. As an example we take
\begin{equation}
S_M=-\ft12 M^2 g^{t/2}\PV{X}^2
       +\ft12 M g\epsilon _{\alpha \beta } \PV{\xi}^\alpha \PV{\xi}^\beta
       +\ft12 M\PV{b}^{\alpha \beta } \PV{b}^{\gamma \delta }
          g_{\alpha \gamma }\epsilon _{\beta \delta }\ .\label{SPVM2dgr}
\end{equation}
To illustrate the arbitrariness in the choice of mass term,
which will also play a role in section~6,
we introduced an arbitrary real parameter $t$ in the mass term for the
scalars. If necessary  one introduces
several copies of PV fields with multiplicities $C_i$,
and imposes the conditions
$\sum_{i}C_i =1$ and $\sum_{i}C_i (M_i)^{2n} = m^{2n}$ for
$n=1,2, ..., n_{max}$.
The divergences of the theory manifest themselves as terms of the form
 $ \sum_{i}C_i (M_i)^{2n} \log M_i^2 $, and have to be absorbed by
 renormalisation before the limit $M_i\rightarrow\infty$ is taken.
 An example will be seen in the next section.
\section{Anomalies}
How anomalies occur in the formal path integral has been explained in
section~(I.4) and the first part of section~II.5.1. Concerning
cohomology let us add that even in the presence of anomalies there is
a nilpotent operator
\begin{equation}
{\cal S}_qF= e^{-\ft i\hbar W} \ft\hbar i \Delta e^{\ft i\hbar W}F=
(W,F)+ \ft\hbar i \Delta F +{\cal A}F\ ,\label{SqF}
\end{equation}
where $W$ is the quantum action $S+\hbar M_1+\ldots $, and $\Delta$
is acting from the left: \footnote{Important properties are
$\Delta FG = \Delta F\cdot G + (-)^FF\,\Delta G +
(-)^F(F,G)$ and $ \Delta ^2 = 0$.}
\begin{equation}
\Delta=(-)^A \frac{\dl}{\partial \Phi^A} \frac{\dl}{\partial
\Phi^*_A}\ .
\end{equation}
The nilpotency of $\Delta$ immediately implies the
nilpotency of ${\cal S}_q$. The anomaly itself is ${\cal A}={\cal S}_q 1$,
and is trivially invariant under ${\cal S}_q$.
If another quantity $M$ can be found such that
${\cal A}= {\cal S}_q M$, then the anomaly can formally be removed by a
redefinition $W'=W+\ft \hbar i \log (1-M)$.
Note however that in general this additional contribution to the action
can not be written as a {\em local} term.
Note also that in $\hbar$ expansions of
expressions like (\ref{SqF}) it has often been
assumed that $\Delta S$ gives only terms of order $\hbar^0$
\cite{BV,anombv}. This may not always be true in a regulated theory.

\label{ss:regul}
In sections I.5.1, I.5.2 and II.5.1 it has been shown why these
formal expressions need regularisation, and how this can be performed
by using PV regulators \cite{anomPV,anombv}. One ends up with a
regulated determination of $\Delta S$ which
depends on the choice of $T$ in the PV mass term of (\ref{SPV}).
(i.e. on the regularisation), on the choice of basis (gauge fixing), ...
but it always takes a value in the same cohomology class: for two such
choices $(a)$ and $(b)$ we have
\begin{equation}
\Delta ^{(b)}S=\Delta ^{(a)}S+ (F,S)\qquad\mbox{with } F\mbox{ local.}
\end{equation}
Rather than repeating the general exposition, we will sketch here
the application to the scalar loops in $d=2$ gravity.

The anomaly at one loop arises from
\begin{equation}
{\cal A}=\ft 1{2\hbar} (S_T,S_T)=\ft 1\hbar(S+S_{PV},S_M)\ .
\end{equation}
We only treat here the  integration over the "matter" PV fields $\PV{X}$,
for the purpose of illustration.
{}From  \eqn{SPVM2dgr} we then only take into account the first term,
and will in particular consider two values for $t$:
\begin{eqnarray} \mbox{at } t=0\ :&&
{\cal A}_0= \frac{M^2}{\hbar}\left[ \PV X^\mu \PV X_\mu
\partial_\alpha \xi^\alpha
-2 \PV X^\mu \PV \xi^\alpha \partial_\alpha X_\mu\right] \nonumber\\
 \mbox{at } t=1\ :&&
{\cal A}_1= \frac{M^2}{\hbar}\sqrt{g}\left[- c   \PV X^\mu \PV X_\mu
 -2 \PV X^\mu \PV \xi^\alpha \partial_\alpha X_\mu\right]\ .
 \end{eqnarray}
After integration over the PV fields the off--diagonal terms
will vanish  and  ($\Box =
\partial_\alpha \sqrt{g} g^{\alpha\beta} \partial_\beta $)
\begin{equation}
\PV X^\mu \PV X_\mu  \rightarrow   \frac{\hbar}{M^2 g^{t/2}- \Box}\ .
\end{equation}
We then use heat kernel techniques. The method was reviewed in
section 5 of (I): one applies (I.47) and (I.50) with
\begin{equation}
\begin{array}{lcc}
t=0\ :& J= \partial_\alpha \xi^\alpha & {\cal R}= \Box \\
t=1\ :& J= -c             &  {\cal R}= \frac{1}{\sqrt{g}}\Box  \ ,
\end{array}
\end{equation}
leading to
\begin{eqnarray}
 {\cal A}_0 &=&\frac{1}{8\pi}\left( M^2 \log M^2 -\ft16 \sqrt{g}\,R+\ft1{12}
\Box \log g\right) (-\partial_\alpha \xi^\alpha) \nonumber\\
 {\cal A}_1 &=&\frac{1}{8\pi}\left( M^2 \log M^2\sqrt{g}
 -\ft16 \sqrt{g}\, R \right) c \ .
\end{eqnarray}
At $t=0$ our mass term is indeed invariant under Weyl
transformations, and not under general $d=2$ coordinate
transformations. At $t=1$ we have the reverse situation.
This explains the occurrence of the Weyl ghosts for $t=1$ and the
diffeomorphism ghosts for $t=0$. Note
that  for $M\to\infty$ there is a diverging part, which is however a
total derivative for $t=0$, and, as will be clear below, is even locally
${\cal S}$--exact for $t=1$. The two expressions for the anomaly
are related by
$
{\cal A}_1-{\cal A}_0 = {\cal S}M_1 $
with
\begin{equation}
M_1=\frac{1}{8\pi}\left( M^2 \log M^2\sqrt{g} -\ft1{12}\log g\,
\sqrt{g}\,R+\ft1{48} \log g \,
\Box \log g\right)\ ,
\end{equation}
showing explicitly that the change in regularisation preserves
the cohomological class of the anomaly.

As already stressed in the previous lectures, the anomaly depends on
$g_{\alpha\beta}=\eta_{\alpha\beta} + b^*_{\alpha\beta}$, i.e. on the
antifields of the gauge--fixed basis.

Knowing that in general the anomaly is an element of the
cohomology of ${\cal S}$, it is interesting to investigate
what the possible classes are, a priori, without
doing the actual anomaly calculation. To this we now turn.
\section{Cohomological analysis of $d=2$ gravity and background charges}
In collaboration with F. Brandt \cite{anom2dgr}, we have performed
an analysis of the cohomology of ${\cal S}$ for local 2--forms
modulo $d$, or $H^g ({\cal S}\mid d)$. The starting point of this analysis
is the symmetry algebra, without assumptions about the classical action.
Technically, we assumed  knowledge of  $S^1$ and $S^2$ only,
i.e. the second and third line of \eqn{S2dgr}.

There is a relation with the local cohomology of 0--forms \cite{Brandt}
\begin{equation} H^g ({\cal S}\mid d) \sim H^{g+2}({\cal S})\ , \end{equation}
using descent equations  ($\omega^g_f$ is an $f$--form of ghost
number $g$)
\begin{equation}
{\cal S}\omega_2^g+d\omega_1^{g+1}=0\ ;
\qquad {\cal S}\omega_1^{g+1}+d\omega_0^{g+2}=0\ ;\qquad
{\cal S}\omega_0^{g+2}=0 \ .
\end{equation}
For this equivalence it is necessary that the reparametrisation ghosts
are present.
The rather general results of \cite{Brandt}, implying e.g. that
derivatives of ghosts are not present in the cohomology,
can not be applied directly to the case of $d=2$
gravity, because they rest on the assumption that
all local symmetries have independent gauge fields, which
is not the case for the Weyl symmetry in our setup.
We will limit ourselves to a description of the results.

We find that a non--trivial cohomology exists for ghost numbers
$g=-1,0,\ldots 4$.
Let us first mention the most general functional with ghost number 0
which does not depend on antifields. This functional can be used as a
classical action, replacing the first line of \eqn{S2dgr}:
\begin{equation}
S_{cl}=\ft12\sqrt{g}\, g^{\alpha\beta}G_{\mu\nu}(X)
\partial_\alpha X^\mu
\cdot \partial_\beta X^\nu
+B_{\mu\nu}(X)\partial_+ X^{[\mu}\cdot \partial_- X^{\nu ]} \ .
\label{genclact}\end{equation}
Here $G_{\mu\nu}(X)$ and $B_{\mu\nu}(X)$ are arbitrary,
but not all physically different:
the following changes relate solutions that are
cohomologically equivalent:
\begin{eqnarray}
G\sim G'_{\mu\nu}=
 G_{\mu\nu}+2\partial_{(\mu}f_{\nu)}-\Gamma_{\mu\nu\rho} f^\rho\ ,
\nonumber\\
B\sim B'_{\mu\nu}=B_{\mu\nu}+2\partial_{[\mu}B_{\nu]}
 +H_{\mu\nu\rho}f^\rho \ ,
\label{transfGB}
\end{eqnarray}
where
\begin{equation}
f_\mu=G_{\mu\nu}f^\nu \ ;\qquad \Gamma_{\mu\nu,\rho}=
\partial_{(\mu}G_{\nu)\rho}-\ft12\partial_\rho G_{\mu\nu}\ ;\qquad
H_{\mu\nu\rho}=3 \partial_{[\mu} B_{\nu\rho]}\ .
\end{equation}
The functions  $B_\mu$ and $f^\mu$ are arbitrary, and the latter
can be interpreted as a target space reparametrisation
$X^\mu\to X^\mu+f^\mu$.

In addition there are antifield--dependent solutions depending on
arbitrary  covariantly constant functions $f^{(\pm)\mu}(X)$,
\begin{equation}
D_\mu^\pm  f^{(\pm)}_\nu \equiv
\partial_\mu f^{(\pm)}_\nu -
\Gamma^\pm_{ \mu\nu,\rho}f^{(\pm)\rho}=0\qquad \mbox{with}\qquad
\Gamma^\pm_{ \mu\nu,\rho}=\Gamma_{\mu\nu,\rho} \pm\ft12
H_{\mu\nu\rho}\ .
\label{pm-connections}
\end{equation}
These solutions are,
using the notations of \eqn{Fsimpl} and $y=h_{++}h_{--}$,
\begin{eqnarray}
M(f^{(+)},f^{(-)})& = & X^*_\mu
\left(\partial_+ \xi^++h_{--}\partial_+\xi^-\right)\cdot
f^{(+)\mu}
- 2\ft1{1-y}\partial_+ h_{--} \cdot  \nabla_+ X^\mu \cdot
f^{(+)} _\mu\  \nonumber \\
&+&X^*_\mu
\left(\partial_- \xi^-+h_{++}\partial_-\xi^+\right)\cdot
 f^{(-)\mu}
- 2\ft1{1-y}\partial_- h_{++} \cdot  \nabla_- X^\mu \cdot
f^{(-)} _\mu\ .
\label{Mhh}
\end{eqnarray}
Before discussing the meaning of these terms,
we give also the solutions for ghost number~1.
There are 2 special solutions that contain no arbitrary
parameters apart from an overall factor:
\begin{equation}
A_{\pm}= c^{\pm}\partial^3_{\pm } h_{\mp \mp }\ .
\end{equation}
Also, there are solutions depending on arbitrary functions
$f^{(\pm )} _{\mu\nu}(X)$:
\begin{equation}
\omega(f^{(+)}_{\mu\nu}, f^{(-)}_{\mu\nu})= \ft{1}{ 1-y} \nabla_+ X^\mu
\cdot \nabla_- X^\nu \cdot \left[ f^{(+)}_{\mu\nu}
\left(\partial_+ \xi^++h_{--}\partial_+\xi^-\right)
+ f^{(-)}_{\mu\nu}
\left(\partial_- \xi^-+h_{++}\partial_-\xi^+\right) \right].
\label{chanomXdep}
\end{equation}
Again, some of these solutions are cohomologically equivalent:
for arbitrary $H^{\pm }_\mu(X)$,
\begin{equation}
f_{\mu\nu}^+\sim f_{\mu\nu}^++ D_\nu^+ H^+_\mu\ ;\qquad
f_{\mu\nu}^-\sim f_{\mu\nu}^-+ D_\mu^- H^-_\nu\ .
\end{equation}
All these solutions of ghost number 1 appear when discussing the
anomalies of general $\sigma$--models with classical action
\eqn{genclact}. In our analysis for the action \eqn{S2dgr} we
met the particular combination
\begin{equation}
A_++A_- \cong \ft12 \, c \,\sqrt{g}\, R
\end{equation}
where $\cong$ means an equality up to ${\cal S}$--exact terms.
In chiral gravity one obtains separately $A_+$ and/or $A_-$ as anomaly.

Now we come to the meaning of the other solutions of ghost number~0,
eq.\eqn{Mhh}. First we give the antibracket\footnote{Note
that $f^{(+)\mu} g^{(+)} _\mu $ is a constant due to
\eqn{pm-connections}, which is valid also for $g$.}
\begin{equation}
\left( M(f^{(+)},f^{(-)}) ,M(g^{(+)},g^{(-)}) \right) \cong -4
f^{(+)\mu} g^{(+)} _\mu A_+  -4 f^{(-)\mu} g^{(-)} _\mu A_-\ .
\label{Mbracket}\end{equation}
This implies that, although $M$ is a local integral with ghost
number zero, it cannot be used as an extra part of the extended
action, as this breaks the master equation. However, the right hand
side of \eqn{Mbracket} allows another interpretation.
If we modify the action by
adding an $M(f^{(+)}, f^{(-)})$--term that is formally of order
$\hbar^{1/2}$, then $(M, M)$ will contribute to the anomaly on the
same level as the one--loop diagrams, modifying the co\"efficient of
the anomaly, and possibly canceling it. This term therefore becomes
an important addition for the quantum theory.
It is the proper generalisation of what is usually called
a "background charge" term. This is apparent
when, for the chiral case, one puts $h_{++} = 0$, drops the
corresponding $\xi ^-$ ghost, and specializes to
$G_{\mu \nu} = -\ft12\delta _{\mu \nu }$. It is then another example
of a term $M_{1/2}$, as treated in section II.5.3,
presented there in the case of chiral $W_3$ to cancel the anomalies
\cite{ROMANS}
(see \cite{hiding,anomw3} for their inclusion in the BV formalism).
For our non--chiral case, again one has to include both chiralities,
and add an appropriate ${\cal S}$--trivial term. This leads to the
dilaton term in the $\sigma$--model.
\section{Conclusions}
In this review we described some of the ongoing work in the
BV quantisation program, with emphasis on the cohomology of
the operator ${\cal S}\equiv (S,\cdot )$. It appears in various
places:
\begin{itemize}
\item the local cohomology at ghost number 0 gives the classical physical
observables.\vspace{-3mm}
\item the cohomology of integrals at ghost number 0 gives the possible
actions, and is also important for counterterms in the renormalisation
procedure \cite{Anselmi}. \vspace{-3mm}
\item the cohomology of integrals at ghost number 1 gives the possible
anomalies. \vspace{-3mm}
\item Antifield dependent anomalies at ghost number 0 correspond to
background charges, possibly allowing a cancellation of anomalies.
\end{itemize}

\noindent{\large\bf Acknowledgments}
\vglue 0.4cm

This work is carried
out in the framework of the European Community Research Programme
"Gauge theories, applied supersymmetry and quantum gravity", with a
financial contribution under contract SC1-CT92-0789.
\vglue 0.5cm
\pagebreak[1]

\end{document}